\documentclass[twocolumn]{aastex63}

\usepackage{amsmath,amssymb,lineno}
\usepackage{newtxmath}
\usepackage{comment}

\submitjournal{ApJ}

\shorttitle{}
\shortauthors{}

\newcommand{\dif}{\mathrm{d}}

\begin{document}

\title{Diffuse flux of ultra-high energy photons from  cosmic-ray interactions in the disk of the Galaxy and implications for the search for decaying super-heavy dark matter}
\correspondingauthor{Zo\'e Torr\`es}
\email{torres@lpsc.in2p3.fr}

\author[0000-0001-8294-6294]{Corinne B\'erat}
\affiliation{Laboratoire de Physique Subatomique \& Cosmologie,
CNRS/IN2P3, Universit\'{e} Grenoble Alpes, Grenoble, France}

\author[0000-0001-6841-3280]{Carla Bleve}
\affiliation{Laboratoire de Physique Subatomique \& Cosmologie,
CNRS/IN2P3, Universit\'{e} Grenoble Alpes, Grenoble, France}

\author[0000-0001-6863-6572]{Olivier Deligny}
\affiliation{Laboratoire de Physique des 2 Infinis Ir\`ene Joliot-Curie,
CNRS/IN2P3, Universit\'{e} Paris-Saclay, Orsay, France}

\author[0000-0001-9787-596X]{Fran\c{c}ois Montanet}
\affiliation{Laboratoire de Physique Subatomique \& Cosmologie,
CNRS/IN2P3, Universit\'{e} Grenoble Alpes, Grenoble, France}

\author[0000-0001-7670-554X]{Pierpaolo Savina}
\affiliation{Laboratoire de Physique des 2 Infinis Ir\`ene Joliot-Curie,
CNRS/IN2P3, Universit\'{e} Paris-Saclay, Orsay, France}
\affiliation{University of Wisconsin-Madison, Department of Physics and WIPAC, Madison, WI, U.S.A}

\author[0000-0002-8327-8459]{Zo\'e Torr\`es}
\affiliation{Laboratoire de Physique Subatomique \& Cosmologie,
CNRS/IN2P3, Universit\'{e} Grenoble Alpes, Grenoble, France}

\begin{abstract}
An estimate of the expected photon flux above $10^{17}~$eV from the interactions of ultra-high energy cosmic rays with the matter in the Galactic disk is presented. Uncertainties arising from the distribution of the gas in the disk, the absolute level of the cosmic ray flux, and the composition of the cosmic rays are taken into account. Within these uncertainties, the integrated photon flux above $10^{17}~$eV is, averaged out over Galactic latitude less than $5^\circ$, between $\simeq 3.2{\times}10^{-2}~$km$^{-2}~$yr$^{-1}~$sr$^{-1}$ and $\simeq 8.7{\times}10^{-2}~$km$^{-2}~$yr$^{-1}~$sr$^{-1}$. The all-sky average value amounts to $\simeq 1.1{\times}10^{-2}~$km$^{-2}~$yr$^{-1}~$sr$^{-1}$ above $10^{17}~$eV and decreases roughly as $E^{-2}$, making this diffuse flux the dominant one from cosmic-ray interactions for energy thresholds between  $10^{17}~$eV and $10^{18}$~eV. Compared to the current sensitivities of detection techniques, a gain between two and three orders of magnitude in exposure is required for a detection below $\simeq 10^{18}$~eV. The implications for searches for photon fluxes from the Galactic center that would be indicative of the decay of super-heavy dark matter particles are discussed, as the photon flux presented in this study can be considered as a floor below which other signals would be overwhelmed.
\end{abstract}


%\linenumbers

\section{Introduction} 
\label{sec:intro}

Cosmic-ray electrons and protons in the $1$--$100$~GeV energy range can produce gamma rays subsequent to their interactions with the interstellar matter in the Galactic disk via high-energy electron Bremsstrahlung, nucleon-nucleon processes, and, to a lesser extent, inverse-Compton interactions with low-energy photons. The diffuse flux of ${>}100~$MeV and GeV gamma rays observed in a narrow band along the Galactic plane~\citep{Narayanan:2008pg,1972ApJ...171...31F,1975ApJ...198..163F,1980NYASA.336..211M,1982A&A...105..164M,Hunter:1997qec,Fermi-LAT:2012edv} is consistent with these expectations~\citep{1984A&A...134...13F,1984ApJ...279..136B,1989ARA&A..27..469B}. 
At higher energies, the flux of cosmic-ray electrons is overwhelmed by that of protons and heavier nuclei, so the dominant photon-producing process is expected to be the nucleon-nucleon collision. The Milagro experiment has allowed the observation of a diffuse flux at TeV energies from the Cygnus region in the Galactic disk~\citep{Milagro:2005xqq}, while the ARGO-YBJ observatory has allowed that of a diffuse flux extended between $25^\circ$ and $100^\circ$ in Galactic longitude for $|b|<5^\circ$ in Galactic latitude~\citep{ARGO-YBJ:2015cpa}. More recently, a similar diffuse flux at sub-PeV energies has been reported from data collected at the Tibet-AS$\gamma$ observatory~\citep{TibetASgamma:2021tpz}. These observations are indeed in agreement with expectations of photon emission from matter irradiated by cosmic rays, see e.g.~\cite{1993APh.....1..281B,1995ApJ...454..774C,2008NIMPA.588...22E,2017JCAP...03..013C,Lipari:2018gzn}. They could shed light on the spectra of cosmic rays in distant regions of the Galaxy~\citep{Lipari:2018gzn}.

Above $10^{17}~$eV, several campaigns are underway to search for ultra-high energy (UHE) photons. On the one hand, the detection of a diffuse flux would allow probing the expectations related to $\pi^0$ decays following the interactions of ultra-high energy cosmic rays \mbox{(UHECRs)} with the background photon fields or dust permeating the source environments or the extragalactic space. On the other hand, the detection of localized fluxes would allow the discovery of cosmic ray sources or, if they are in the direction of the Galactic center, to highlight the presence of decaying super-heavy dark matter produced in the early Universe. However, the interstellar matter of the Galactic disk is irradiated by the UHECRs. Therefore, a diffuse flux of UHE photons is expected to be produced by the same mechanism responsible for the emission in the GeV--PeV range. An estimate of this flux, in addition to that mentioned above originating from the interaction of UHECRs with the photon fields permeating the Universe, is needed to know the ``background'' that could hide the emission from sources in the Galactic disk or the Galactic center. 

Because a few percent only of UHECRs interact at least once with the matter in the Galactic disk, and because the Galaxy is essentially transparent to photons with energies in excess of $10^{17}~$eV (see Appendix~\ref{sec:app1}), estimations of the diffuse flux (per steradian) of UHE photons at energy $E$, $\phi_\gamma(E,\mathbf{n})$, can be made in a straightforward manner by integrating the position-dependent emission rate per unit volume and unit energy along the line of sight,
\begin{equation}
    \label{eqn:photonflux}
    \phi_\gamma(E,\mathbf{n})=\frac{1}{4\pi}\int_0^\infty \dif s~q_\gamma(E,\mathbf{x}_\sun+s\mathbf{n}).
\end{equation}
Here, $\mathbf{x}_\sun$ is the position of the Solar system in the Galaxy and $\mathbf{n}\equiv\mathbf{n}(\ell,b)$ is a unit vector on the sphere pointing to the longitude $\ell$ and latitude $b$, in Galactic coordinates. The high degree of isotropy of UHECRs above $10^{17}~$eV, as observed on Earth~\citep{PierreAuger:2020fbi}, leads to an equally isotropic irradiation of gas distribution in the Galactic disk, and therefore an isotropic emission that explains the $1/4\pi$ factor. The emission of UHE photons stems from the creation and decay of unstable mesons, mostly $\pi^0$ but also, to a lower extent, $\eta$ and $\eta'$, in the inelastic interactions of UHECRs with interstellar gas,
\begin{equation}
    \label{eqn:photonsource}
    q_\gamma(E,\mathbf{x})=4\pi\sum_{i,j}n_j(\mathbf{x})\int_E^\infty\dif E'\phi_i(E')\sigma_{ij}(E')\frac{\dif N_{ij}^\gamma}{\dif E}(E',E).
\end{equation}
The summation over $j$ accounts for the different gas elements (mainly molecular and atomic hydrogen, and helium), while that over $i$ accounts for the different UHECR-projectile elements. The $n_j(\mathbf{x})$ function stands for the position-dependent number density of the $j$th element, $\phi_i(E')$ is the energy-differential flux per steradian (energy spectrum) of the $i$th UHECR element with energy $E'$, $\sigma_{ij}(E')$ is the inelastic cross section of the collision between the projectile $i$ and the target $j$, and $\dif N_{ij}^\gamma/\dif E(E',E)$ is the inclusive spectrum of photons generated in the $ij$ collision subsequently to the decay of all unstable particles created in the collision. The integration is carried out over all UHECR energies $E'>E$ that allow for generating photons with energy $E$. Finally, the $4\pi$ factor results from the integration of the (isotropic) UHECR flux over solid angle.

The aim of this paper is thus to estimate $\phi_\gamma(E,\mathbf{n})$ by means of Equations~\ref{eqn:photonflux} and~\ref{eqn:photonsource}. To do so, the UHECR flux shaped by the various contributing mass components is described in Section~\ref{sec:xgal}; a survey of the up-to-date modeling of the interstellar gas density in the disk of the Milky Way is presented in Section~\ref{sec:galdisk}; and the photon production in cosmic ray-gas collisions is detailed in Section~\ref{sec:collisions}. The  diffuse flux of UHE photons is then obtained in Section~\ref{sec:flux} for various energy thresholds, before implications for the searches for super-heavy dark matter decaying into photons are drawn in Section~\ref{sec:shdm}. Finally, the significance of the results is summarized in Section~\ref{sec:summary}.

\section{Ultra-high energy cosmic-ray flux} 
\label{sec:xgal}

\begin{figure}[t]
\centering
\includegraphics[width=\columnwidth]{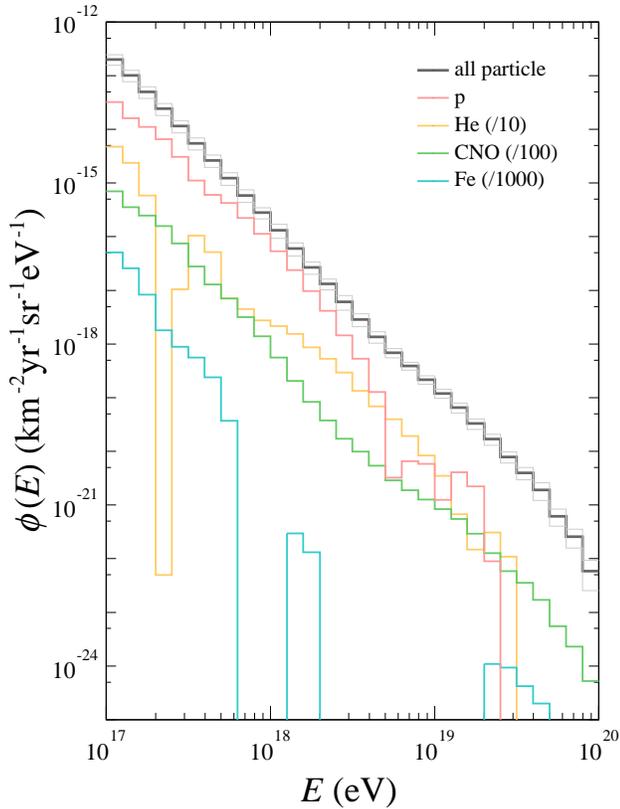}
\caption{All-particle spectrum of UHECRs in thick grey, as reported in~\citep{PierreAuger:2021epjc}. The various mass-discriminated energy spectra are obtained by making use of the Sibyll2.3 model of hadronic interactions to interpret the longitudinal profiles of extensive air showers observed at the Pierre Auger Observatory~\citep{Aab:2014aea,Bellido:2017cgf}.}
\label{fig:CRflux}
\end{figure}

In the energy range of interest, knowledge of the energy spectrum and mass composition of UHECRs is possible only through indirect measurements, via the extensive air showers that they produce in the atmosphere. In particular, the energy-dependent mass composition is accessible on a statistical basis only. Hence, the various $\phi_i(E)$ components must be inferred by combining the all-particle energy spectrum of UHECRs, $\phi(E)=\sum_i\phi_i(E)$, and the abundances of the different elements as a function of energy. 

The all-particle energy spectrum of UHECRs $\phi(E)$ has been measured between $10^{17}~$and $10^{18}~$eV from several dedicated contemporary experiments~\citep{knurenko2013cosmic,Bertaina:2015fnz,Abbasi:2018xsn,IceCube:2019hmk,Budnev:2020oad}. At higher energy, above $10^{18}~$eV, the fall of the flux required a leap in cumulative exposure to collect an increased influx of events~\citep{PierreAuger:2015eyc,TA2012}. The measurement of the spectrum carried out at the Pierre Auger Observatory above $10^{17}~$eV, reported in~\citep{PierreAuger:2021epjc}, is the one relying on the largest cumulated exposure with a single detector type, avoiding the necessity to combine measurements, the matching of which inevitably implies additional systematic effects. In addition, the energy scale is set calorimetrically via fluorescence telescopes, avoiding assumptions on the mass of the particles. We use this measurement in the following, shown as the thick grey line in Fig.~\ref{fig:CRflux}, and make an explicit use of the reported covariance matrix $\mathbf{\sigma_\phi}$ that accounts for both statistical and systematic uncertainties to calculate (see Appendix~\ref{sec:app2}) the 2-sided 16\% quantiles of the underlying distribution of spectrum estimator, denoted as $\phi_+(E)$ and $\phi_-(E)$ (shown as the thin grey lines in Fig.~\ref{fig:CRflux}). This will allow us to propagate in Section~\ref{sec:flux} the uncertainties in the spectrum measurement, coming for the most part from those in the energy scale, in the calculation of the photon flux. 

The approximately power-law shape of the spectrum in $E^{-3}$ hides beneath a complex intertwining of different astrophysical phenomena, which might be exposed by looking at the spectrum of different primary elements. An illustration of the contributions of protons, He, N and Fe nuclei to the total spectrum is shown, by scaling the He (N) [Fe] components by 10 (100) [1000] so as to distinguish each of them. They are obtained from the abundances of elements reported in~\cite{Aab:2014aea,Bellido:2017cgf}, by weighting the total spectrum with the fractions of the corresponding mass group as inferred from the distributions of the depths of the shower maximum, $X_\text{max}$, measured at the Pierre Auger Observatory and modelled according to, for this specific illustration (Fig.~\ref{fig:CRflux}), the hadronic model Sibyll2.3~\citep{riehn2017hadronic}. A linear interpolation (in $\log_{10}{E}$) is applied to smooth the fluxes at the bin-center values. One can observe the steep fall-off of the Fe component above $10^{17}$\,eV, which is along the lines of the long-standing scenario for Galactic CRs characterised by a  rigidity-dependent maximum acceleration energy~\citep{1961NCim...22..800P}. On the other hand, one can also observe the presence of protons, He, and N nuclei, at least up to $\simeq 10^{19}$\,eV, at which point the proton flux begins to quench and that of the N hardens. This is consistent with a rigidity-dependent maximum acceleration energy scenario for extra-galactic sources, see e.g.~\cite{Aloisio:2013hya,Taylor:2015rla,Aab:2016zth,Aab:2020rhr}. Within such a scenario, the presence of intermediate He and N nuclei in the energy range between $\simeq 10^{17}$\,and $\simeq 10^{18.7}$\,eV, where the Galactic Fe is declining and presumably the extra-galactic protons are rising, could be fueled by a second Galactic ``B component'', as first suggested in~\cite{Hillas:2005cs}. 

With caution over the interpretation of the $X_\text{max}$ measurements, we thus use the data described above to model the various $\phi_i(E)$ components. As a proxy of the estimation of the uncertainties, we use the three different results reported in~\cite{Bellido:2017cgf} for the fractions of mass groups inferred from three different hadronic models, namely EPOS-LHC~\citep{Pierog:2013ria}, \mbox{QGSJetII-04}~\citep{ostapchenko2013qgsjet} and Sibyll2.3. The systematic differences are on the order of, or larger than, the statistical uncertainties, which, in the absence of an available covariance matrix governing the correlations between energy bins, are not accounted for in this study.

Our final goal is to calculate photon fluxes above $10^{17}$\,eV, which are mainly produced by cosmic rays of energy larger than $A~10^{18}$\,eV, with $A$ the atomic number of the particles. Hence mainly extra-galactic UHECRs, the flux of which is assumed to illuminate the Galaxy uniformly in the relevant energy range, can contribute to the photon fluxes sought for. Subsequently, the various $\phi_i(E)$ components are assumed to fill homogeneously the Galactic volume. This neglects flux-amplification effects preferentially in the disk that would be indicative of larger residence times for particles experiencing stronger values of the Galactic magnetic field. This is reasonable as numerical simulations of particle trajectories in the field show that the residence time of cosmic rays tends to equate their free escape time for particle rigidities above $\simeq 1~$EV~\citep{1998AstL...24..139Z,Kaapa:2021}.

\section{Interstellar gas density in the Milky Way} 
\label{sec:galdisk}

The interstellar medium in the Galaxy is known to provide a mass of a few $10^9~M_\sun$, which is distributed predominantly in the disk of the Milky Way. Hydrogen ($\simeq 90\%$) and helium ($\simeq 10\%$) in gaseous state make up most of this mass, while a small fraction is contributed by silicate and graphite dust. Depending on its temperature, three forms of the hydrogen gas are at play: atomic (\mbox{H\,{\sc i}}), molecular (H$_2$), and ionized (\mbox{H\,{\sc ii}}) hydrogen, while helium remains mostly neutral due to its much higher first ionization potential. In this study, we use two models of the spatial distribution of the gas recently developed from existing data. The differences in the final photon emission from UHECR-gas interactions between the two models are interpreted as contributing to the systematic uncertainties of  $\phi_\gamma(E,\mathbf{n})$. The first model, developed by~\cite{Lipari:2018gzn} and dubbed as model A hereafter, aims at capturing the large-scale properties of the gas based on an axially and up-down symmetric distribution with a scale height increasing with the radial distance from the Galactic center, hence without attempting to describe accurately fine details. It is worth noting the diffuse gamma-ray fluxes observed around GeV energies and sub-PeV ones could be reproduced using this model. The second model, developed by~\cite{Johannesson:2018bit} and dubbed as model B, incorporates in addition spiral arms and accounts for the warping of the disk so as to constrain models of cosmic-ray propagation and to trace the structure of the Galaxy. We describe below the main features of the models, without reproducing the detailed parameterizations available in the original papers. A comprehensive description of the data used to construct these models can be found in, e.g.,~\cite{Ferriere:2001rg}.

The \mbox{H\,{\sc i}} component represents a large fraction of the mass. The most important probe of its distribution is the line at 21 cm observed in emission or absorption. 
%The absorption coefficient is determined by the small difference in the population between the singlet and triplet states, which in turn is governed by the excitation temperature (spin temperature) of the emitting gas. The spin temperature, strongly correlated to the temperature of the gas, varies from a few tens of K (cold neutral medium) to a few thousands of K (warm neutral medium). 
The density is inferred to be constant in the distance range from 4 to $\simeq 10~$kpc from the Galactic center and decreases steadily at larger distances, but falls exponentially in the vertical direction with a scale length of $\simeq 250$~pc for the warm component and $\simeq 130$~pc for the cold one. The density of the cold component is  highly non-uniform, as illustrated by its filling factor of the volume of the thin disk less than 0.1.

Although almost equally abundant in the interstellar medium, the distribution of H$_2$ is less well known. This is because such a homonuclear diatomic molecule does not possess a permanent electric dipole moment, and therefore has no simple rotational transition. It can be however observed in the UV region through electronic transitions, and indirectly inferred from the rotational transition of the $^{12}$C$^{16}$O molecule (CO hereafter for convenience), which is mostly excited through collisions with H$_2$ and which is the second most abundant molecule in the interstellar medium. The integrated intensity of the CO lines turns out to be almost linearly related to the column density of H$_2$ so that the CO observations at radio wavelengths nicely complement the surveys of the 21-cm line, the proportionality factor being taken as $2{\times}10^{20}~$cm$^{-2}~$(K~km~s$^{-1}$)$^{-1}~$\citep{Bolatto:2013ks}. 

\begin{figure}[t]
\centering
\includegraphics[width=\columnwidth]{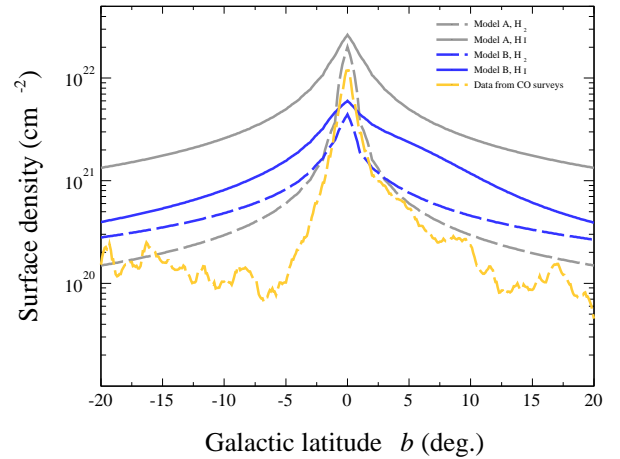}
\caption{Surface density average over Galactic longitude as a function of Galactic latitude for neutral (\mbox{H\,{\sc i}}) and molecular (H$_2$) hydrogen, as modeled by~\cite{Lipari:2018gzn} (model A) and by~\cite{Johannesson:2018bit} (model B). The surface density of H$_2$ as obtained from the composite survey of~\cite{Dame:2000sp} is shown as the orange line.}
\label{fig:hdens}
\end{figure}

The surface densities for neutral and molecular hydrogen, as inferred from the models A and B and averaged over Galactic longitude, are shown in Fig.~\ref{fig:hdens}. For reference, the surface density of H$_2$ as obtained from the composite survey of~\cite{Dame:2000sp} is shown as the orange line. Once summing the \mbox{H\,{\sc i}} and H$_2$ components, the two models give rise to some differences by a factor of a few in units of $10^{22}~$cm$^{-2}$. We consider these differences as sources of systematic uncertainties in the photon flux calculation.

The distribution of the helium contribution, $\simeq 10\%$ of that of hydrogen, is assumed to follow closely the hydrogen one. No additional gas element is considered in this study. In particular, the small contribution from ionized hydrogen is neglected.

\section{Photon production in cosmic ray-gas collisions} 
\label{sec:collisions}

\begin{figure}[t]
\centering
\includegraphics[width=\columnwidth]{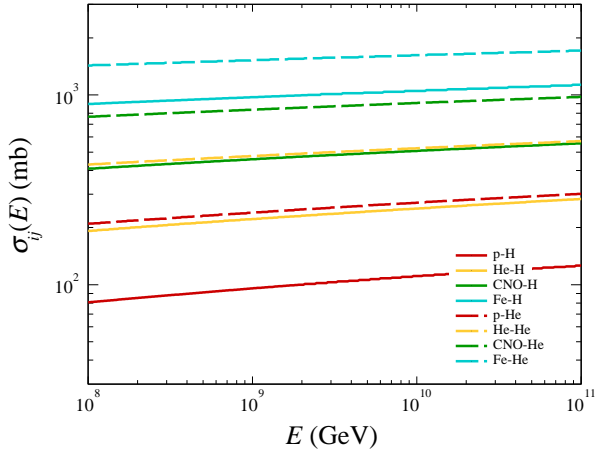}
\caption{Inelastic cross section as a function of energy for the various processes of interest.}
\label{fig:sigma}
\end{figure}

\begin{figure}[t]
\centering
\includegraphics[width=\columnwidth]{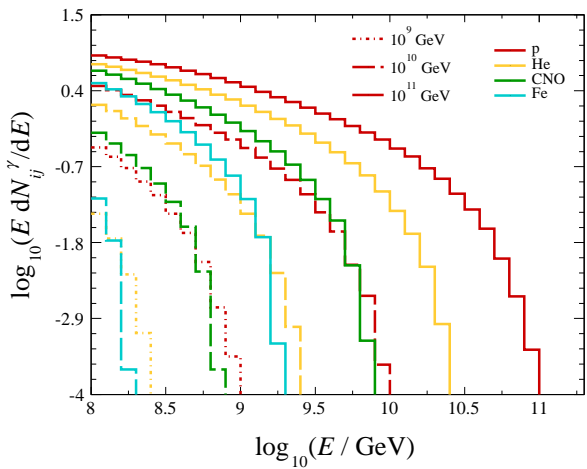}
\caption{Photon yield per UHECR interaction for different primaries and incoming energies.}
\label{fig:yield}
\end{figure}

Cosmic rays interact with the interstellar medium through the different processes described in the introduction. These collisions create neutral pions $\pi^0$ whose most probable mode of decay produces two photons:
\begin{equation}
    \label{decaypi0}
    \pi^0 \rightarrow 2\gamma.
\end{equation}
The production of secondary particles such as $K$, $\rho$, $\eta$ decaying into neutral pions also occurs during these interactions, resulting in an additional flux of photons.

To obtain the relevant inelastic cross sections $\sigma_{ij}(E')$ as well as the inclusive spectrum of photons $\dif N^{\gamma}_{ij}/\dif E(E',E)$, which corresponds to the mean number of photons in the energy range $[E,E+\dif E]$ produced in a single interaction between the cosmic ray element $i$ with energy $E'$ and the gas element $j$ at rest, we use Cosmic Ray Monte Carlo (CRMC)~\citep{crmcsource}. This package provides different cosmic-ray event generators that model the production of secondary particles occurring in hadronic interactions. It also allows the follow-up of the decays of the secondary particles that are likely to produce photons. In this study, the hadronic interaction model EPOS-LHC~\citep{Pierog:2013ria} is chosen. 

Thus, for different values of the UHECR energies ranging from $10^{17}$ eV to $10^{20}$ eV (in steps of \mbox{$\Delta\log_{10}(E')=0.5$}), $100,000$ collisions are simulated for each couple of cosmic ray and target element $(i,j)$ of interest here. The spectra of the photon energies are then produced for each cosmic-ray energy, with a bin width of \mbox{$\Delta\log_{10}(E)=0.05$}. 

The inelastic cross sections $\sigma_{ij}(E)$ obtained with CRMC are shown in Fig.~\ref{fig:sigma} for each cosmic-ray element as continuous lines for H targets and dashed ones for He targets. They range typically in the hundred of millibarns, reaching thousand ones for the heaviest collisions. The photon yields are shown in Fig.~\ref{fig:yield} for the different primaries and three cosmic-ray energies, $10^9~$GeV (dotted), $10^{10}~$GeV (dashed) and $10^{11}~$GeV (continuous). For a fixed photon energy $E<E'$, the expected increase of the yield with the incident energy of cosmic rays is observed. On the other hand, for a fixed cosmic-ray energy, the yield is also observed to  depend strongly on the cosmic-ray mass. Given that a single nucleus generally intervenes in each interaction, this behavior is expected from the energy available in each nucleus, which, compared to the total energy of the nucleus, is reduced by the atomic number of the nucleus. As a result, the photon fluxes are thus more copiously produced at higher energies by the lightest primaries.

\section{Diffuse flux of ultra-high energy photons} 
\label{sec:flux}

\begin{figure}[t]
\centering
\includegraphics[width=\columnwidth]{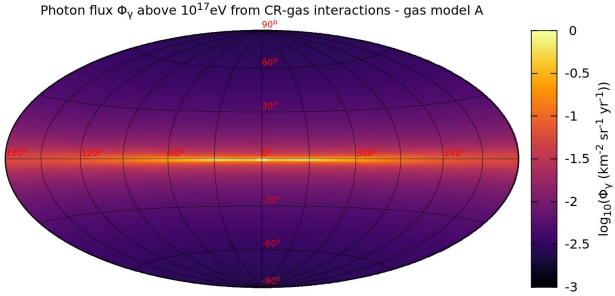}
\caption{Photon flux in Galactic coordinates (Hammer projection) expected from UHECR-gas interactions, integrated above $10^{17}~$eV. Model A is used for the gas distribution in the Galactic disk.}
\label{fig:mapA}
\end{figure}

\begin{figure}[t]
\centering
\includegraphics[width=\columnwidth]{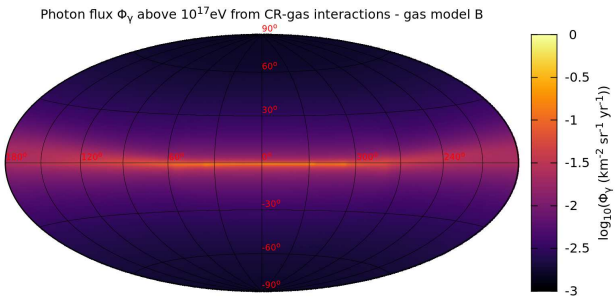}
\caption{Same as Fig.~\ref{fig:mapA}, using model B for the gas distribution in the Galactic disk.}
\label{fig:mapB}
\end{figure}

On inserting the various ingredients detailed in Sections~\ref{sec:xgal},~\ref{sec:galdisk}, and~\ref{sec:collisions} into equation~\ref{eqn:photonsource} and subsequently equation~\ref{eqn:photonflux}, the photon flux per energy unit sought for is obtained. In Fig.~\ref{fig:mapA} and Fig.~\ref{fig:mapB} are shown the results by considering the two models of gas distribution in the Galactic disk described in Section~\ref{sec:galdisk}, and by integrating over energy above $10^{17}~$eV. As expected, the flux is concentrated a few degrees around the Galactic plane, amounting to $\simeq 8.7{\times}10^{-2}$~km$^{-2}$yr$^{-1}$sr$^{-1}$ once averaged out over $|b|\leq5^\circ$ in the case of model A ($\simeq 3.2{\times}10^{-2}$~km$^{-2}$yr$^{-1}$sr$^{-1}$ in the case of model B). The pattern is brighter in the innermost region of the disk for both models.

The energy dependence of the photon flux is shaped by that of the cosmic-ray flux, inheriting from its main features but shifted about a decade earlier. Averaged over a 5$^\circ$-band around the Galactic plane, the photon flux is about $10^{-5}$ that of the UHECRs for energy thresholds ranging from $10^{17}~$eV to $10^{18}~$eV, before steepening to $10^{-6}$ above $10^{19}~$eV and dropping sharply at higher thresholds.

\begin{figure}[t]
\centering
\includegraphics[width=\columnwidth]{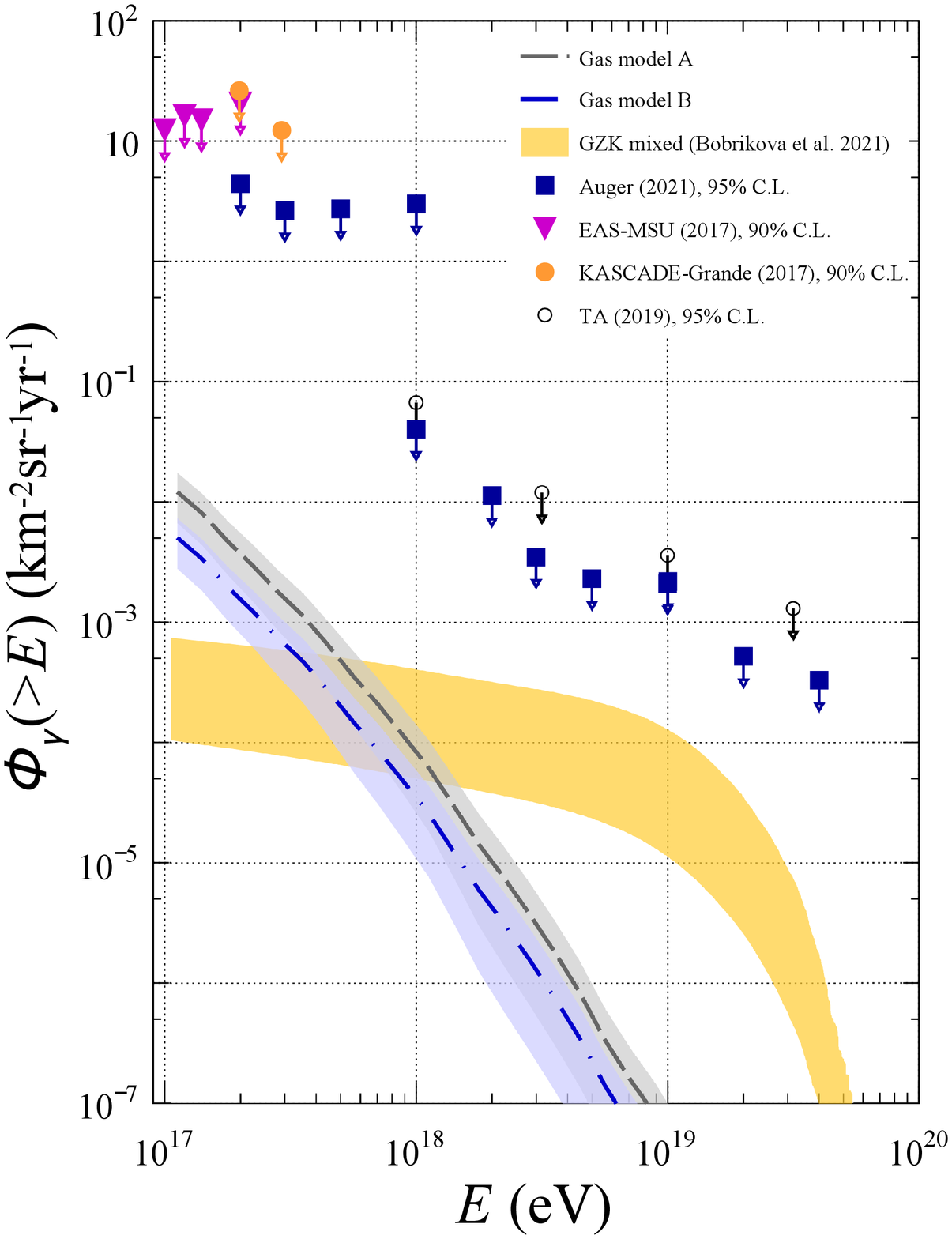}
\caption{Upper limits on diffuse photon fluxes reported by~\cite{Fomin:2017ypo} (pink triangles, EAS-MSU),~\cite{Apel:2017ocm} (filled orange circles, KASCADE-Grande),~\cite{Abbasi:2018ywn} (open circles, Telescope Array), and~\cite{Savina:2021zrn} (dark blue squares, Auger) for several energy thresholds. Expected fluxes from UHECRs interactions with the gas distribution in the Galaxy are shown as the grey dashed line (model A) and grey dashed-dotted line (model B), as well as from UHECRs interactions with background photon fields (GZK mixed) estimated by~\cite{Bobrikova:2021kuj}.}
\label{fig:phi-vs-UL}
\end{figure}

The calculated photon flux can be compared to several existing upper limits above $10^{17}~$eV from several observatories. Shown in Fig.~\ref{fig:phi-vs-UL} as pink triangles~\citep{Fomin:2017ypo}, filled orange circles~\citep{Apel:2017ocm}, dark blue squares~\citep{Savina:2021zrn}, and open circles~\citep{Abbasi:2018ywn}, they appear to be between $\simeq 2.5$ and $\simeq 3$ orders of magnitude above our expectations for gas distribution models A and B for energy thresholds between $10^{17}~$eV (amounting to $\Phi_\gamma\simeq 1.1{\times}10^{-2}~$km$^{-2}~$yr$^{-1}~$sr$^{-1}$ for model A) and $\simeq 10^{18.5}~$eV ($\Phi_\gamma\simeq 2.0{\times}10^{-6}~$km$^{-2}~$yr$^{-1}~$sr$^{-1}$), and even higher for larger thresholds. The bands correspond to the systematic uncertainties arising from those in the all-particle energy spectrum and those of the hadronic interaction model used to infer the mass composition of UHECRs. Note that to be compared to the upper limits on the diffuse fluxes illuminating the whole field of view of the  observatories considered, the expected fluxes A and B reported in Fig.~\ref{fig:phi-vs-UL} are ``calibrated'' on the whole sphere as follows:
\begin{equation}
    \Phi_\gamma(E)=\frac{1}{4\pi}\int_E^\infty\dif E'\int_{4\pi}\dif\mathbf{n}~\phi_\gamma(E',\mathbf{n}).
\end{equation}
For comparison with other astrophysical expectations, another photon flux, resulting from interactions of extragalactic UHECRs with the background photon fields permeating the Universe, is shown as the orange band~\citep{Bobrikova:2021kuj}. The hadrons that cause the creation of $\pi^0$ mesons, which then decay into photons, may be primary proton cosmic rays, or secondary ones produced by the photo-disintegration of nuclei interacting inelastically with a cosmic background photon. Since the nucleons produced inherit the energy of the fragmented nucleus divided by its atomic number, this photon flux therefore depends on the nature of the UHECRs. Other dependencies come from the maximum acceleration energy of the nuclei at the sources and the shape of the energy spectrum of the accelerated particles. Different assumptions about these ingredients have led to different estimates in the literature, see e.g.~\cite{Kampert:2011hkm}. The recent estimate reported in~\cite{Bobrikova:2021kuj} that considers matching the energy spectrum measured at the Pierre Auger Observatory simultaneously with the mass composition inferred from these data is considered here. Considered as systematic uncertainties for the expected photon flux, they give the orange band in Fig.~\ref{fig:phi-vs-UL}. 

Interestingly, among the two photon fluxes produced by cosmic-ray interactions, the one explored in this study appears to be dominant in the energy range between $10^{17}~$and $\simeq 10^{18}~$eV. Its observational signature, that of an excess of a few degrees around the Galactic plane, makes it unambiguous to identify once the required sensitivity is reached in the future. At higher energy thresholds, the other one, expected to be isotropic, takes over. 

\section{Implications for searches for super-heavy dark matter}
\label{sec:shdm}

Between $10^{17}~$and $\simeq 10^{18}~$eV, the flux of UHE photons reported in this study can be seen as a floor that prevents probing sources in the Galactic disk or patterns of super-heavy dark matter (SHDM) decaying primarily from the Galactic center and producing a photon flux $\phi_\gamma^{\mathrm{DM}}$ below the level of $\phi_\gamma$. This is because the detection of such an overwhelmed flux would require an exposure much larger than $1/\phi_\gamma^{\mathrm{DM}}$ so as to pick up a significant excess above $\phi_\gamma$ integrated within some angular and energy ranges. The production of super-heavy particles in the early Universe indeed remains a possible solution to the dark matter puzzle because of the high value of the instability energy scale in the Standard Model of particle physics, which, according to current measurements of the Higgs boson mass and the Yukawa coupling of the top quark, ranges between $10^{10}$ and $10^{12}~$GeV~\citep{Degrassi:2012ry,Bednyakov:2015sca}. The Standard Model can therefore be extrapolated without encountering inconsistencies that would make the electroweak vacuum unstable up to such energy scales (and even to much higher ones given the slow evolution of the instability scale up to the Planck mass~\citep{Degrassi:2012ry}) where new physics could arise, giving rise to a mass spectrum of super-heavy particles that could have been produced during post-inflation reheating by various mechanisms, see e.g.~\cite{Ellis:1990iu,Ellis:1990nb,PhysRevLett.79.4302,Chung:1998zb,Garny:2015sjg,Ellis:2015jpg,Dudas:2017rpa,Kaneta:2019zgw,Mambrini:2021zpp}. The existence of a photon-flux floor in the direction of the Galactic center has important implications for SHDM searches using UHE photons from the Galactic center, and translates, depending on their mass $M_X$, into a ``ceiling'' for the exploration of the particle lifetime that we now discuss.

\begin{figure}[t]
\centering
\includegraphics[width=\columnwidth]{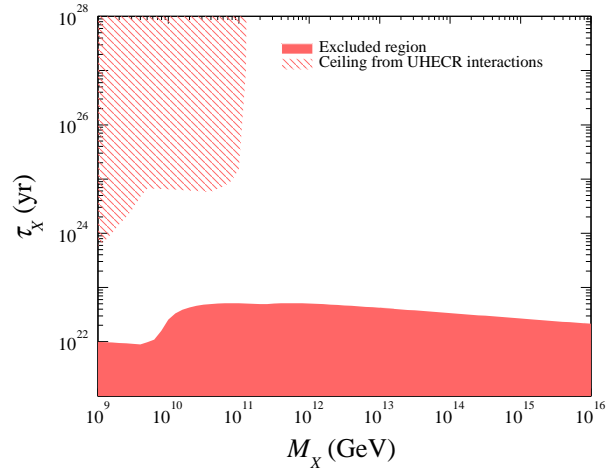}
\caption{Allowed region of mass and lifetime of super-heavy dark matter particles decaying into standard-model ones. The filled red region is excluded from the upper limits in UHE photon fluxes while the hatched one corresponds to the observational ceiling discussed in the text.}
\label{fig:m-tau}
\end{figure}

Secondary UHE photons produced by the decay of SHDM particles can be observed today if these particles have a long enough lifetime. This usually requires a fine tuning between the coupling constant of the particle with those produced and the dimension of the operator governing the decay~\citep{PhysRevD.67.125019}. Alternatively, SHDM particles can be protected from decay in the perturbative domain by the conservation of new quantum number(s) but disintegrate by instanton non-perturbative effects~\citep{Kuzmin:1997jua}. This mechanism offers the possibility of providing metastable particles with lifetime $\tau_X$, which can produce detectable secondaries such as nucleons and photons. In particular, the flux (per steradian) of photons produced can then be obtained as the sum of a  contribution from the Galaxy and of an isotropic extragalactic one -- see e.g.~\cite{Berezinsky:1998rp,Benson:1999ie,Medina-Tanco:1999ktg,Aloisio:2006yi,Siffert:2007zza,Kalashev:2016cre,Alcantara:2019sco}. The Galactic contribution, which is the dominating one, is obtained by integrating the position-dependent emission rate per unit volume and unit energy along the path of the photons in the direction $\mathbf{n}$:
\begin{equation}
\label{eqn:phiDM}
\phi_\gamma^{\mathrm{DM}}(E,\mathbf{n})=\frac{1}{4\pi M_X\tau_X}\frac{\dif N}{\dif E}\int_0^\infty\dif s\rho_{\mathrm{DM}}(\mathbf{x}_\sun+s\mathbf{n}).
\end{equation}

Here, $\rho_{\mathrm{DM}}$ is the energy-density profile of dark matter in the Galaxy. There are uncertainties in the determination of this profile~\citep{Guepin:2021ljb}. However, the typical $1$-degree angular resolution of UHECR-Observatory capabilities cannot probe the uncertainties of the parametric models~\cite{Alcantara:2019sco}. We thus assume here the traditional Navarro-Frenk-White profile~\citep{Navarro:1995iw}, normalized so that $\rho(\mathbf{x}_\sun)=0.3~$GeV~ cm$^{-3}$ in the solar system. The other term, $\dif N/\dif E$, stands for the energy spectrum of the secondary photons, which depends on the hadronization process. Various computational schemes, see e.g.~\cite{Aloisio:2004ap}, show that the spectrum of the final state particles (photons, nucleons, and neutrinos) varies as $E^{-1.9}$. We use this widely-agreed energy dependence, normalizing the spectrum in the same way as in~\cite{Alcantara:2019sco}

From the limits on $J_{\gamma}(E)$, constraints can be inferred in the plane $(\tau_X,M_X)$ by requiring the all-direction flux calculated by averaging Eqn.~\ref{eqn:phiDM} over all directions to be less than the limits. For a specific upper limit, a scan of the value of the mass $M_X$ is carried out so as to infer a lower limit of the $\tau_X$ parameter. This defines a curve. By repeating the procedure for each upper limit on $J_\gamma(E)$, a set of curves is obtained, reflecting the sensitivity of a specific energy threshold to some range of mass. The union of the excluded region is shown as the filled red region in Fig.~\ref{fig:m-tau}, which is similar to that of previous studies~\citep{Kalashev:2016cre,Ishiwata:2019aet}.

Although the directional dependence of $\phi_\gamma(E,\mathbf{n})$ (Eqn.~\ref{eqn:photonflux}) is different from that of $\phi_\gamma^{\mathrm{DM}}(E,\mathbf{n})$ (a hot spot falling steeply a few degrees away from the Galactic center), it shall mask the presence of decaying SHDM for values of $\phi_\gamma^{\mathrm{DM}}(E,\mathbf{n})$ in the direction of the Galactic center below that of $\phi_\gamma(E,\mathbf{n})$ in the same direction. 
By scanning $M_X$, we can also determine the corresponding value of $\tau_X$ beyond which the expected values of $\phi_\gamma^{\mathrm{DM}}(E)$ would be overwhelmed by the $\phi_\gamma(E)$ averaged-directional fluxes. This results in the ceiling region shown as the hatched region in Fig.~\ref{fig:m-tau}. The ceiling is observed to affect mass values up $M_X\simeq 10^{11}~$GeV, reflecting the cut-off of the photon fluxes produced by cosmic-ray interactions. Below $M_X\simeq 10^{10}~$GeV, the ceiling region is most constrained by the photon flux inferred in this study. For larger masses, on the other hand, the constraints obtained from the photon flux inferred in~\cite{Bobrikova:2021kuj} take over. The sharp increase just above $M_X=10^{11}~$GeV reflects the sharp drop of the photon fluxes originating from cosmic-ray interactions for energy thresholds above a few times $10^{19}~$eV, leaving a large region of the phase space unaffected by the ceiling. \\

\section{Summary}
\label{sec:summary}

The all-sky flux a few degrees around the Galactic plane of UHE photons produced by UHECR interactions in the disk of the Galaxy amounts to, integrated above $10^{17}~$eV, $1.1{\times}10^{-2}~$km$^{-2}~$yr$^{-1}~$sr$^{-1}$. It turns out to be the dominating one from cosmic-ray interactions for energies between $10^{17}~$and $\simeq 10^{18}~$eV. Although out of reach of the current and planned observatories, it may be responsible for the first detection of photons in this energy range once future experiments reach the required exposure, unless brighter sources of photons are discovered while cumulating this exposure. This flux will also be a floor below which other signals coming from the same direction would be overwhelmed. This is relevant for searches for photon fluxes from the Galactic center that would be indicative of the decay of super-heavy dark matter particles, and this translates into a ceiling region for the lifetime of the putative particles to be explored below $M_X\simeq10^{11}~$GeV. 

Equally expected from UHECR-gas interactions, UHE fluxes of neutrinos are produced by charged pions or neutrons decays. Quantitative estimates are left for a future study.

%\acknowledgments

\section*{Acknowledgments}
We thank Roger Clay for his comments. This work was made possible by with the support of the Institut Pascal at Universit\'e Paris-Saclay during the Paris-Saclay Astroparticle Symposium 2021, with the support of the P2IO Laboratory of Excellence (program ``Investissements d’avenir'' ANR-11-IDEX-0003-01 Paris-Saclay and ANR-10-LABX-0038), the P2I axis of the Graduate School Physics of Universit\'e Paris-Saclay, as well as IJCLab, CEA, IPhT, APPEC, the IN2P3 master projet UCMN and EuCAPT ANR-11-IDEX-0003-01 Paris-Saclay and ANR-10-LABX-0038).
 
\newpage
\appendix

\section{UHE photon absorption in the Galactic disk}
\label{sec:app1}
The radiation length of UHE photons propagating in the matter of the Galactic disk is of order $100~$g~cm$^{-2}$, much larger than the column density that is traversed, which is at most a few tenths of g~cm$^{-2}$. UHE photon absorption by the dust is thus negligible. The most important process is pair production $\gamma\gamma\rightarrow e^+e^-$ with the radiation fields that permeate the Galaxy. In the energy range considered in this study ($\geq 10^{17}~$eV), the radiation field that dominates by far is that of the cosmic microwave background~\citep{Lipari:2018gzn}. In this case, because the density of the radiation field is uniform in space, the optical depth for photons observed on Earth with energy $E$ in the direction $\mathbf{n}$, $\tau(E,\mathbf{n},x)$, is proportional to the travelled distance $x$:
\begin{equation}
    \tau(E,\mathbf{n},x)=K(E,-\mathbf{n})x.
\end{equation}
The function $K(E,\mathbf{n})$ is the interaction probability per unit length, obtained by integrating over the energy and the (isotropic) angular distribution of the target photons: 
\begin{equation}
    K(E,\mathbf{n})=\int \dif\mathbf{\kappa} (1-\cos{\theta_{\gamma\gamma}})n_{\gamma}(\epsilon)\sigma_{\gamma\gamma}(s).
\end{equation}
In this expression, $\mathbf{\kappa}$ is the 3-momentum of the target photon and $\epsilon=|\mathbf{\kappa}|$ its energy, $\theta_{\gamma\gamma}$ is the angle between the interacting photons, and $\sigma_{\gamma\gamma}(s)$ is the pair-production cross section expressed as a function of the square of the center-of-mass energy $s=2E\epsilon(1-\cos{\theta_{\gamma\gamma}})$, which, in the high-energy regime that is the relevant one here, behaves, modulo a logarithmic correction, as $1/s$ (and thus as $1/E$):
\begin{equation}
    \sigma_{\gamma\gamma}(s)\simeq \frac{3\sigma_{\mathrm{T}}m_{\mathrm{e}}^2c^4}{2s}\left(\ln{\frac{s}{m_{\mathrm{e}}^2c^4}}-1\right),
\end{equation}
with $\sigma_{\mathrm{T}}$ the Thomson cross section. In this way, the optical depth averaged over angle behaves roughly as $\tau(E)\simeq 0.03/E$ (in kpc$^{-1}$). Plugging an absorption factor $\exp{(-\tau(E,\mathbf{n}))}$ into the integrand in equation~\ref{eqn:photonflux} leads however to small reductions, of order of 10\% at $10^{17}~$eV and much smaller at higher energies. This is much less than the various systematic uncertainties considered in the study. 

\section{Propagation of the uncertainties in the all-particle energy spectrum}
\label{sec:app2}

Denoting as a vector the set of measurements of the energy spectrum in each energy bin, $\boldsymbol{\phi}=\{\phi_1,\phi_2,...,\phi_N\}$, the $\boldsymbol{\phi}_+$ vector is defined as the set of values that satisfy
\begin{equation}
\label{eqn:quantile}
    \frac{1}{\sqrt{(2\pi)^N\mathrm{det}\boldsymbol{\sigma_\phi}}}\int_{\boldsymbol{\phi}_+}^\infty\dif\boldsymbol{\phi}\exp{\left(-\frac{1}{2}\delta\boldsymbol{\phi}^\mathrm{T}\boldsymbol{\sigma^{-1}_\phi}\delta\boldsymbol{\phi}\right)}=0.84.
\end{equation}
The notation $\delta\boldsymbol{\phi}$ stands for a random fluctuation around the set of observed values. To solve equation~\ref{eqn:quantile} for the unknown $\boldsymbol{\phi}_+$ and the corresponding one for $\boldsymbol{\phi}_-$, we build the probability distribution function of $\boldsymbol{\phi}$ by ``whitening'', or ``standardizing'' the covariance matrix in the following way. We first decompose the covariance matrix as $\boldsymbol{\sigma_\phi}=\boldsymbol{R}\boldsymbol{\sigma_\phi'}\boldsymbol{R}^\mathrm{T}$, where the columns of the rotation matrix $\boldsymbol{R}$ are the eigenvectors of $\boldsymbol{\sigma_\phi}$ and $\boldsymbol{\sigma_\phi'}=\mathrm{diag}(\sigma_1'^2,\dots,\sigma_N'^2)$, with $\sigma_i'^2$ the eigenvalues of $\boldsymbol{\sigma_\phi}$. Then, it is straightforward to build a transformed flux $\delta\boldsymbol{\phi}'$ for which the covariance matrix is the identity one by considering the re-scaled flux $\delta\boldsymbol{\phi}'=\boldsymbol{\sigma_\phi'}^{-1/2}\boldsymbol{R}^\mathrm{T}\delta\boldsymbol{\phi}$. In this way, the directional properties of the initial covariance matrix are eliminated in the new coordinate system so that random realizations of $\delta\boldsymbol{\phi}'$ can be drawn from a $N$D normal distribution, with the set of values $(\delta\phi'_1,\dots,\delta\phi'_N)$ independent from each other. The corresponding values for $\delta\boldsymbol{\phi}$ are recovered from the transformation $\delta\boldsymbol{\phi}=\boldsymbol{R}\boldsymbol{\sigma_\phi'}^{1/2}\delta\boldsymbol{\phi}'$. By repeating a large number of times the procedure, the 2-sided 16\% quantiles defining $\boldsymbol{\phi}_+$ and $\boldsymbol{\phi}_-$ are finally estimated.  

\bibliographystyle{aasjournal.bst}
\bibliography{biblio}

\end{document}